\documentclass[prb,twocolumn,showpacs,superscriptaddress,preprintnumbers,amssymb]{revtex4-2}
\usepackage{graphicx}
\usepackage{color}
\usepackage{dcolumn}
\usepackage{bm}
\usepackage{graphicx}
\usepackage{amsmath}
\usepackage{hyperref}
\usepackage{comment}
\usepackage{slashed}

\newcommand{\beq}{\begin{equation}}
\newcommand{\eeq}{\end{equation}}
\newcommand{\beqn}{\begin{eqnarray}}
\newcommand{\eeqn}{\end{eqnarray}}

\newcommand{\ua}{\uparrow}
\newcommand{\da}{\downarrow}
\newcommand{\ra}{\rightarrow}

\newcommand{\cC}{ {\cal C} }

\newcommand{\cL}{ {\cal L} }

\newcommand{\cS}{ {\cal S} }

\newcommand{\cZ}{ {\cal Z} }

\newcommand{\ii}{\mathrm{i}}

\newcommand{\llangle}{\langle\!\langle}
\newcommand{\rrangle}{\rangle\!\rangle}

\newcommand{\SU}{\mathrm{SU}}
\newcommand{\U}{\mathrm{U}}

\newcommand{\tr}{\mathrm{tr}}

\definecolor{orange_custom}{rgb}{0.93, 0.47, 0.2}

\begin{document}

\title{Inequality for Strong-Weak Spontaneous Symmetry Breaking in Fermionic Open Quantum systems}

\author{Abhijat Sarma}

\author{Cenke Xu}

\affiliation{Department of Physics, University of California, Santa Barbara, CA 93106, USA}


\begin{abstract}

Under decoherence, an initial Gaussian (free-fermion) state evolves into a non-Gaussian mixed state, so the resulting decohered fermionic state is not exactly solvable in general. We show through an inequality that a class of R\'{e}nyi-2 correlators of the decohered fermion state are upper-bounded by the R\'{e}nyi-2 correlator serving as a proximate diagnostic of strong-weak spontaneous symmetry breaking (SW-SSB) of the charge-U(1) symmetry. This inequality holds for arbitrary decoherence strength and suggests that decoherence drives fermionic quantum matter toward U(1) SW-SSB. We also make connections between our inequality and other subjects such as projected quantum spin Hall insulator and Dirac spin liquid states. 

\end{abstract}

\maketitle

\section{Introduction}

A quantum system loses its quantum characteristics and becomes more classical through interacting and entangling with environment, a process called decoherence. It is natural to ask whether decoherence is a smooth evolution, or there can be a sharp transition. When one looks for a sharp transition in physics, the most unambiguous organizing principle is the framework of symmetry: if the symmetry of the system changes under decoherence, e.g. there is a spontaneous symmetry breaking, then a sharp transition must exist. In recent years, it has been gradually understood that the process of decoherence indeed often involves a sharp transition called the strong-weak spontaneous symmetry breaking (SW-SSB)~\cite{lee2023, lessa2025,weinstein2025,gu2025,sala2024, huang2025, zhang2025b, kim2024, chen2025, sa2025, ziereis2025, zerba2025, hauser2025}, marking the onset of classical physics, such as the emergence of classical hydrodynamics~\cite{huang2025,hauser2026}. 

An open quantum system has a strong symmetry $G$ when there is a ``doubled" conservation of quantum numbers of $G$ in the system and the environment separately. In the Lindbladian description of the decoherence process, strong symmetry means the jump operators commute with the charge of $G$, and the quantum number in the ket and bra spaces of the density matrix are separately conserved~\cite{buca2012,degroot2022}. For example, when the system decoheres through a density-density system-environment interaction, a process called dephasing, the particle number of the system and environment are separately conserved, and in the density matrix formalism the particle numbers in the ket (left) and bra (right) spaces are separately conserved, i.e. there is a strong-$\U(1)$ symmetry. 
The effect of dephasing can be represented conveniently in the ``doubled space". By taking transpose of the bra space, any density matrix, pure or mixed, is mapped to a pure quantum state in the doubled Hilbert space: \beqn \rho = \sum_{ij} \rho_{ij} |i\rangle \langle j| \ \ra \ |\rho \rrangle = \sum_{ij} \rho_{ij} |i\rangle \otimes |j\rangle. \eeqn This seemingly trivial manipulation provides a powerful and intuitive picture for dephasing and SW-SSB. Starting with an initial quantum state, under dephasing its density matrix in the doubled space reads: \beqn |\psi(g)\rrangle &\sim&  e^{ - H_{\rm eff}(g)}| \psi(0)\rrangle, \cr \cr  H_{{\rm eff}}(g) &=& \sum_i \frac{g}{2} (n_{i,L} - n_{i,R})^2, \label{doubleH}\eeqn i.e. the process of dephasing is mapped to an imaginary-time evolution of the initial doubled state, with an ``interlayer" attractive interaction between the left and right spaces identified as a bilayer system. Then it is natural to expect that, starting with an initial fermion state, the effective interlayer attractive interaction may lead to an interlayer Cooper pair condensate, which is precisely the doubled space representation of the SW-SSB, as the Cooper pair correlation in the doubled space is nothing but the R\'{e}nyi-2 correlator~\cite{su2025}, the (proximate) diagnostic for the SW-SSB. 

But since dephasing is mapped to interaction in the doubled space, the system is in general not a Gaussian state, even if the initial state is Gaussian (free-fermion). Therefore the exact solution for a dephased fermion system is usually not available, and one often needs to resort to numerical methods. In this work, we will prove that, though the exact solution is unknown, dephasing does drive the system toward SW-SSB, since a family of R\'{e}nyi-2 correlators are upper bounded by the correlator $\cC_\ast$ that serves as the proximate diagnostic for the SW-SSB. 

We note here that the logic of the bound is similar to the known Weingarten inequality proven for a seemingly completely unrelated problem, the Quantum chromodynamics
(QCD)~\cite{weingarten}. This inequality showed that the pseudoscalar pion meson should have the strongest correlation, or equivalently the smallest meson mass in QCD, under certain general assumptions.

\section{General Tight-binding model}

Let us consider a state $|\Psi_0\rangle$ of $N$-flavors of degenerate fermions, and it is the ground state of the following tight-binding Hamiltonian: \beqn H_0 = \sum_{a = 1}^N \sum_{ij} c^\dagger_{a,i} h_{ij} c_{a,j}. \eeqn We do not assume any symmetry of $h_{ij}$, including translation, meaning the tight-binding model can have disorder. The doubled space representation of the system reads \beqn |\Psi(0) \rrangle = |\Psi_0 \rangle_L \otimes |\Psi_0^\ast \rangle_R. \eeqn 
In the doubled space the generator of strong-U(1) symmetry is $Q_s = \sum_i n_{i,L} + n_{i,R}$.
For convenience, let us perform a particle-hole transformation for the right-space: \beqn && c_{i, \ua} = c_{i,L}, \ \ \ c_{i,\da} = c^\dagger_{i,R}, \cr\cr && n_{i,\ua} = n_{i,L}, \ \ \ n_{i,\da} = N - n_{i,R}. \label{PHR} \eeqn After transformation Eq.~\ref{PHR}, the strong-U(1) charge becomes the pseudo-spin density, and the weak-U(1) generator becomes the total charge density of $c$: \beqn Q_s = \sum_{a,i} \, c^\dagger_{a,i} \sigma^3 c_{a,i}, \ \ \ Q_w = \sum_{a,i} \, c^\dagger_{a,i} c_{a,i} \eeqn
The original tight-binding parent Hamiltonian for the right-space state $|\Psi_0^\ast \rrangle_R$ is just $h^\ast_{ij}$, then after transformation Eq.~\ref{PHR} it becomes $  - h^\ast_{ji} = - h_{ij}$. We have used the hermiticity of $h_{ij}$, i.e. $h_{ij} = h^\ast_{ji}$. Hence after the transformation Eq.~\ref{PHR}, the parent Hamiltonian for the doubled state $|\Psi(0)\rrangle $ reads: \beqn H^d_0 &=& \sum_{a,i} c^\dagger_{a,i} \left( h_{ij} \otimes \sigma^3 \right) c_{a,j}. \eeqn 

The ``effective interaction" $H_{\rm eff}(g)$ in Eq.~\ref{doubleH} now becomes \beqn H_{\rm eff}(g) = \sum_i \frac{g}{2} (n_{i,\ua} + n_{i,\da})^2 + {\rm Const}. \label{Heff} \eeqn We have used the fact that $|\Psi(0)\rrangle$ has a fixed total particle number, therefore $\sum_i n_{i,\ua} + n_{i,\da}$ is a constant. 

\begin{figure}
\centering
\includegraphics[width=0.8\linewidth]{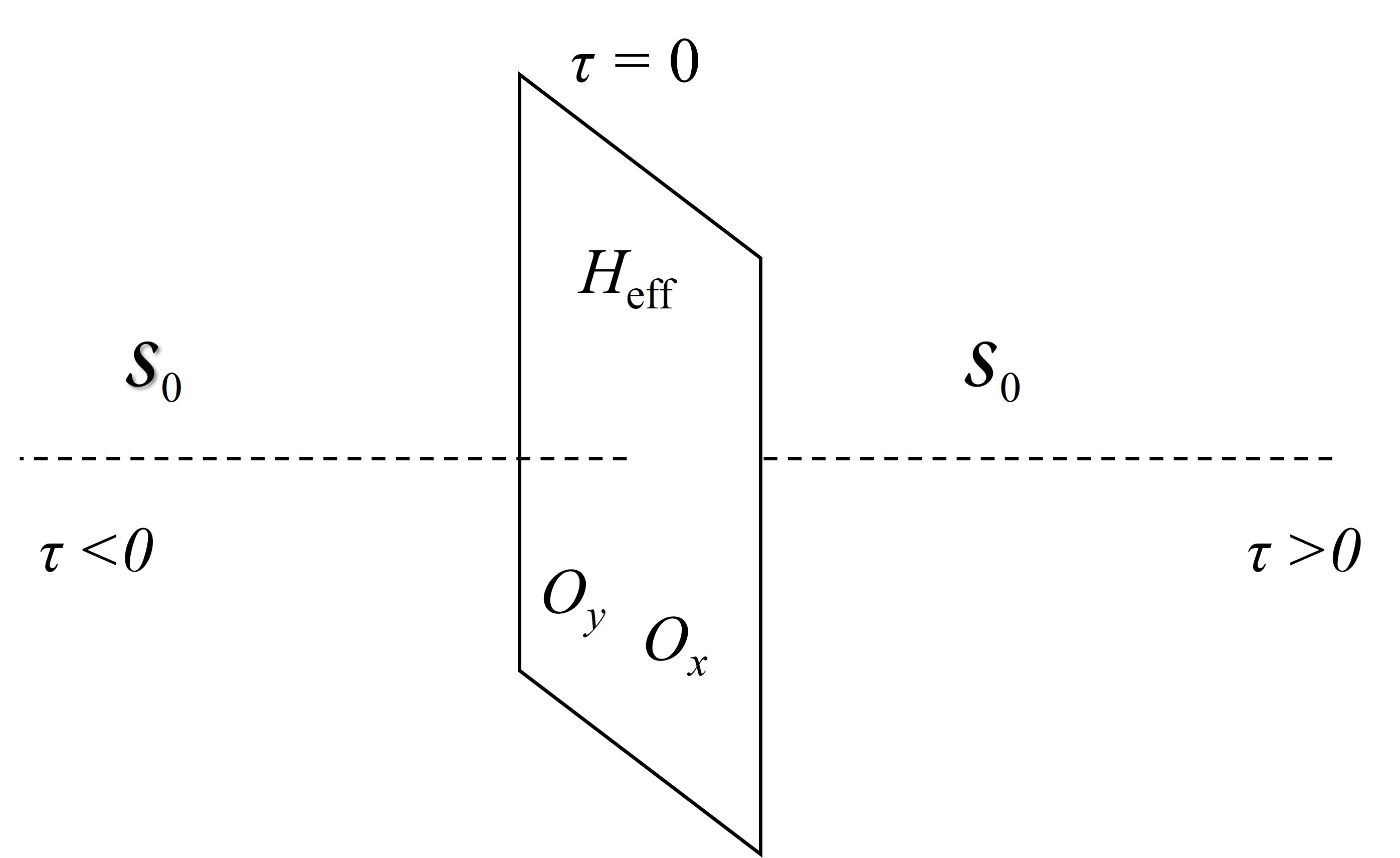}
\caption{The Euclidean spacetime path integral (Eq.~\ref{PI0}) representation of correlator $C(x, y)$. Action $\cS_0$ is defined in the entire spacetime; the operators $O_x$ and $O_y$ as well as the effective Hamiltonian $H_{\rm eff} (g)$ arising from dephasing are inserted at $\tau = 0$. }
\label{path}
\end{figure}

We aim to evaluate the correlation function 
\beqn C(x, y) = \frac{ \llangle \Psi(g)| O_x \, O_y |\Psi(g)\rrangle }{\llangle \Psi(g)| \Psi(g)\rrangle}, \eeqn which maps to the following form in the Euclidean space-time path-integral: \beqn C(x, y) &=& \frac{1}{\cZ} \int D c D c^\ast \ O_{x,\tau = 0} \, O_{y, \tau=0} \cr\cr &\times& \exp\left( - \cS_0 - 2 H_{{\rm eff}}(g)_{\tau = 0} \right), \cr\cr \cS_0 &=& \int d\tau \sum_{ij} \, c^\ast_i \partial_\tau c_i + c_i^\ast \left( h_{ij} \sigma^3 \right) c_j \label{PI0} \eeqn Note that $\cS_0$ resides in the entire spacetime to generate wave function $|\Psi(0)\rrangle$, and $H_{\rm eff}$ is only inserted at the temporal slab $\tau = 0$ (Fig.~\ref{path}). Similar path-integral representation as well as the mapping to temporal defect have been explored before~\cite{garratt2023,Lee2025symmetryprotected,lee2023}.

We first consider the simple {\bf onsite} fermion-bilinear operator $O_x = c^\dagger_x \sigma^\mu \Omega c_x$. $\Omega$ is a $N \times N$ Hermitian matrix in the flavor space. Here $\mu= 0 - 3$, and $\sigma^0$ is just the identity matrix in the pseudo-spin space. In particular, the operator ${\cal O} = c^\dagger \sigma^1 \Omega c$ is a Cooper pair operator before transformation Eq.~\ref{PHR}: ${\cal O} = c^t_R  \Omega c_L + h.c.$. Our goal is to prove that the correlator $C(x, y)$ of $O$ is upper-bounded by the correlator ${\cal C}_\ast(x, y)$ of ${\cal O}$.

To proceed we introduce the Hubbard–Stratonovich (HS) field $\phi_{x}$ in the path-integral, and the partition function becomes \beqn  \cZ &=& \int D c Dc^\ast D \phi \ e^{- \cS}. \cr\cr \cS &=& \cS_0  + \sum_i \left( \ii 2 \phi_{i} c^\ast_i c_i  + \frac{1}{g} \phi_i^2 \right)_{\tau = 0}. \eeqn 
The correlator $C(x, y)$ can be evaluated following the procedure of determinant quantum Monte Carlo. We first evaluate the correlator for each configuration of the bosonic field $\phi_x$, then sum over all configurations of $\phi_x$: \beqn C(x, y) &=& - \frac{1}{\cZ} \int D \phi {\cal M}[\phi] \, \tr \left( \sigma^\mu \Omega S_{x,y} \sigma^\mu \Omega S_{y,x} \right)_\phi  \cr\cr {\cal M}[\phi] &=& ({\rm det} D_\phi )^{N} \exp \left( - \frac{1}{g} \sum_i \ \phi_i^2 \right). \label{C2} \eeqn Here ${\cal M}[\phi]$ is the measure of path-integral of $\phi_x$, and operator $D_\phi$ is \beqn D_\phi = \partial_\tau + h_{ij} \sigma^3 + \ii 2 \phi \delta(\tau), \ \ \ S = D^{-1}_{\phi}.  \eeqn 
The operator $D_\phi$ satisfies a key algebra: \beqn \sigma^1 D_\phi \sigma^1 = - D_\phi^\dagger, \label{algebra1} \eeqn which used the fact that $\partial_\tau$ and $\ii \phi \delta(\tau)$ are both {\it anti-Hermitian} and {\it commute} with $\sigma^1$, while $h_{ij} \sigma^3$ is {\it Hermitian} and {\it anti-commutes} with $\sigma^1$. 
For each configuration $\phi_x$, the amplitude of correlator $|C_{\phi}(x, y)|$ is bounded by the following inequality \beqn && |C_{\phi}(x, y)| = |\tr\left( \sigma^\mu \Omega S_{x, y} \sigma^\mu \Omega S_{y, x} \right)_\phi| \cr\cr &\leq& \sqrt{ \tr \left( \sigma^\mu \Omega S_{x, y} S_{x, y}^\dagger \Omega \sigma^\mu \right)_\phi } \sqrt{ \tr \left( \sigma^\mu \Omega S_{y, x} S_{y, x}^\dagger \Omega \sigma^\mu  \right)_\phi } \cr\cr &=& \sqrt{ \tr \left( \Omega S_{x, y} S_{x, y}^\dagger \Omega \right)_\phi } \sqrt{ \tr \left( \Omega S_{y, x} S_{y, x}^\dagger \Omega \right)_\phi } \cr\cr &=& - \tr \left( \sigma^1 \Omega S_{x, y} \sigma^1 \Omega S_{y, x} \right)_\phi = {\cal C}_{\ast,\phi}(x, y). \label{inequality1} \eeqn We have used the Cauchy-Schwarz inequality, \beqn |\tr\left( U V \right)| \leq \sqrt{\tr \left( U U^\dagger \right) } \sqrt{\tr \left( V V^\dagger \right) }. \eeqn Also, since $\Omega$ is a Hermitian matrix in the flavor space, and $S$ is flavor-blind, $\Omega$ commutes with $S$. Thirdly, we used the algebra Eq.~\ref{algebra1}, $\sigma^1 S_{x,y} \sigma^1 = - S^\dagger_{y,x}$, $\sigma^1 S_{y,x} \sigma^1 = - S^\dagger_{x,y}$, and cyclicity of trace. The last line of Eq.~\ref{inequality1} is precisely the correlator ${\cal C}_{\ast,\phi}$ of ${\cal O} = c^\dagger \sigma^1 \Omega c$, with the background bosonic configuration $\phi_x$. 

We have proven that, for each configuration of $\phi_x$, the correlator of $O = c^\dagger \sigma^\mu \Omega c$ is upper-bounded by the correlator of ${\cal O}$. To show that this inequality holds after summing over $\phi_x$, we need to prove that $({\rm det} D_\phi)^N$ is non-negative, at least for certain choice of $N$. The algebra Eq.~\ref{algebra1} is again the key for this proof: \beqn {\rm det}[D_\phi] = {\rm det}[\sigma^1 D_\phi \sigma^1] =  {\rm det}[- D_\phi^\dagger] = \left( {\rm det}[D_\phi] \right)^\ast. \label{positive}\eeqn Here we have used the fact that the dimension of matrix $D_\phi$ is even thanks to the pseudo-spin degrees of freedom naturally built in the doubled space representation. Eq.~\ref{positive} implies that ${\rm det}[D_\phi]$ is real, and an even integer $N$ is a sufficient (though not necessary) condition for $\left( {\rm det}[D_\phi] \right)^N$ to be nonnegative. With nonnegative measure ${\cal M}[\phi]$ of $\phi$, the triangle inequality leads to \beqn |C(x,y)| \leq \frac{1}{\cZ} \int D \phi {\cal M[\phi]} \, |C_\phi(x, y)|, \eeqn and Eq.~\ref{inequality1} implies that $|C(x,y)|$ is further bounded by ${\cal C}_\ast(x, y)$.

As we discussed, after transformation Eq.~\ref{PHR} the strong-U(1) generator becomes the pseudo-spin density $Q_s = \sum_i \, c^\dagger_i \sigma^3 c_i$. The operator ${\cal O} = c^\dagger \sigma^1 \Omega c$ transforms into $c^\dagger \sigma^2 \Omega c$ under strong-U(1) as an inplane pseudo-spin order parameter, and it is invariant under the weak-U(1) generated by $Q_w$. Therefore its R\'{e}nyi-2 correlator is a diagnostic of the U(1) SW-SSB. We stress that the R\'{e}nyi-2 correlator is a proximate diagnostic, as the precise diagnostic is the fidelity or R\'{e}nyi-1 correlator~\cite{lessa2025,weinstein2025}. 

We now consider the {\bf non-onsite} fermion-bilinear operator $O_x = c^\dagger_{x} \sigma^\mu \Omega c_{x+\delta}$, and the R\'{e}nyi-2 correlator between $O_x$, $O_y$:
\beqn && |C_{\phi}(x, y)| = |\tr\left( \sigma^\mu \Omega S_{x + \delta, y } \sigma^\mu \Omega S_{y + \delta, x} \right)_\phi| \cr\cr &\leq& 
\sqrt{ \tr \left( \Omega S_{x + \delta, y} S_{x + \delta, y}^\dagger \Omega \right)_\phi } \sqrt{ \tr \left( \Omega S_{y + \delta, x} S_{y + \delta, x}^\dagger \Omega \right)_\phi } \cr\cr &=& \sqrt{{\cal C}_{\ast,\phi}(x + \delta, y)} \sqrt{{\cal C}_{\ast,\phi}( x, y+\delta)}. 
\label{inequality2} \eeqn Though the last line is not exactly ${\cal C}_{\ast,\phi}(x, y)$, if $|x - y| \gg \delta$, it differs only by microscopic shifts of the operator positions. After coarse-graining $x$ and $y$ in the neighborhood with size $\delta$, the microscopic shifts in Eq.~\ref{inequality2} are washed out, so the resulting averaged correlator should have the same long-distance behavior as ${\cal C}_{\ast,\phi}(x, y)$. Then we expect the correlation of non-onsite operators to be also bounded by the asymptotic scaling of ${\cal C}_{\ast}(x, y)$.

\begin{figure}
\centering
\includegraphics[width=\linewidth]{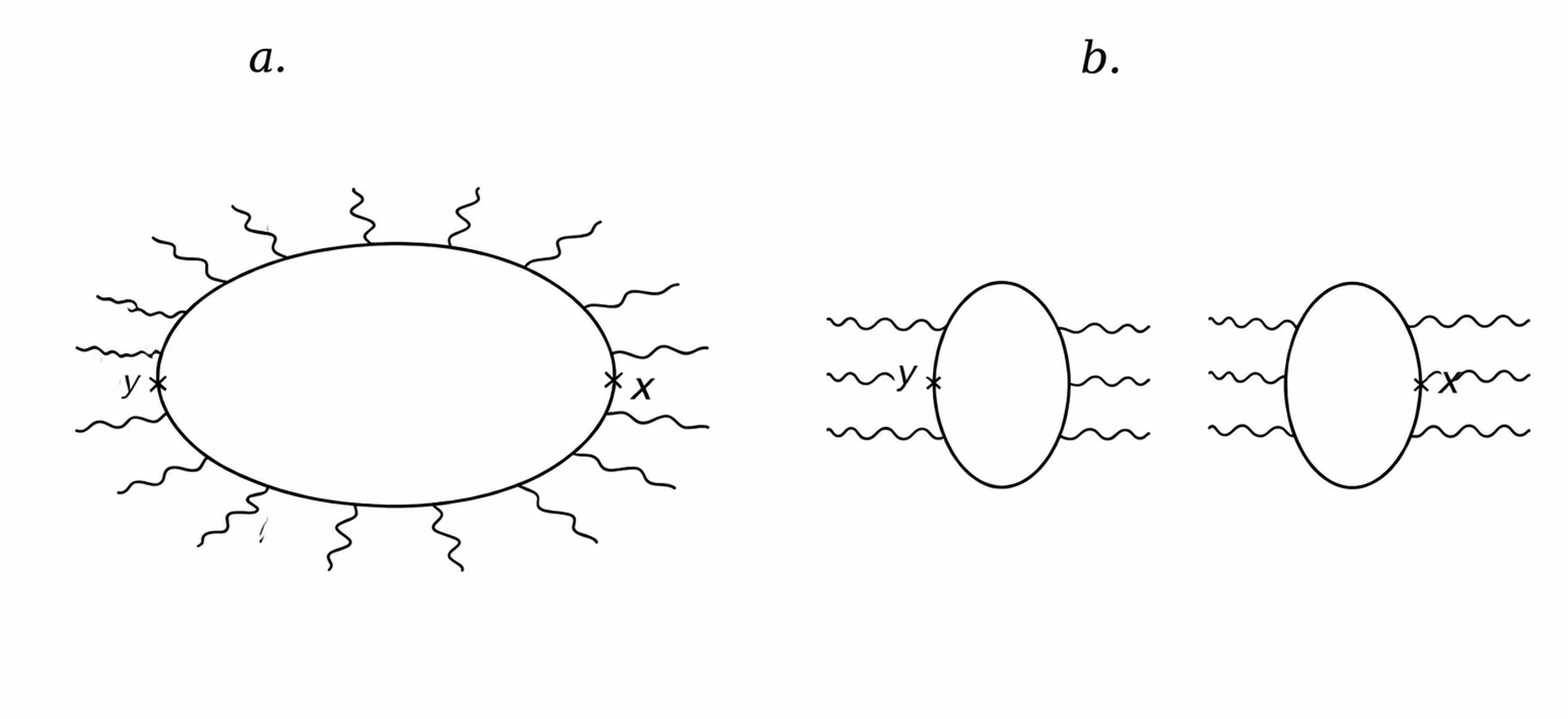}
\caption{Two types of Feynman diagrams (see also Ref.~\cite{vafawitten,witteninequality}) potentially contributing to the correlator of $O = c^\dagger \sigma^\mu \Omega c$. Type-$a$. ``Connected" diagrams, position $x$ and $y$ are connected by Fermion lines; and Type-$b$. ``disconnected" diagrams. The inequality discussed in this work applies to type-$a$ diagrams. Type-$b$ diagram vanishes when $\tr[\Omega] = 0$.  }
\label{diagram}
\end{figure}

An important remark to make is that, the inequality proven here applies to the ``connected" correlator of $O$, meaning that in terms of Feynman diagrams, $O_x$ and $O_y$ are connected by Fermion Green's functions (Fig.~\ref{diagram}, see also Ref.~\cite{witteninequality,vafawitten}).But all the disconnected diagrams involve a factor of $\tr[\Omega]$, therefore to ensure that the disconnected contributions vanish, we simply need to assume that $\Omega$ is a traceless Hermitian matrix in the flavor space, such as the generator of the $\SU(N)$ flavor symmetry. Then for these choices of $\Omega$, the correlator of $O$ is bounded by that of ${\cal O}$. The fact that the R\'{e}nyi-2 correlator of a class of operators are upper-bounded by that of ${\cal O}$, indicates that dephasing drives a quantum system toward SW-SSB. 

\section{Applications}

\subsection{Other Decoherence Channels}

Our inequality can be naturally generalized to scenarios when the jump operator of the Lindbladian is not a simple density operator. For example, let's assume $N = N_f N_c$ with even $N_f$. Now the fermions carry an extra ``color" index running $1, \cdots N_c$. We can consider jump operators \beqn L^I_i = c^\dagger_i T^I c_i, \label{jump}\eeqn where $T^I$ are $N_c \times N_c$ Hermitian matrices in the color space, such as the generators of the $\SU(N_c)$. In this case, the path-integral formalism for the R\'{e}nyi-2 correlator still applies, with \beqn D_{\phi} = {\bf 1}_{N_f} \otimes \left( \partial_\tau + h_{ij} \sigma^3 + \sum_I 2 \ii \phi^I T^I \delta(\tau) \right). \eeqn 
The key results especially the algebra Eq.~\ref{algebra1} still hold, i.e.
the correlators of all the fermion bilinear operators $c^\dagger_x \sigma^\mu \Omega c_{x + \delta}$ are upper-bounded by the correlator of ${\cal O}_x = c^\dagger_x \sigma^1 \Omega c_x$, except now $\Omega$ is a traceless Hermitian $N_f \times N_f$ matrix in the flavor space. 

\subsection{Gutzwiller state and Emergent order}

As was noted in recent work~\cite{su2025}, the doubled state of the fully dephased fermion state is precisely the same as the Gutzwiller projected state, a subject that has been studied broadly in condensed matter systems as a proximate construction for strongly interacting electrons~\cite{hasting2000,hermele2005,ran2007,hermele2008,becca,becca2,hesl,hesl2,lauchlisl}. In particular, if the original state is a Chern insulator, the doubled state becomes a quantum spin Hall state, 
because if $|\Psi_0\rangle_L$ has Chern number $+N$, $|\Psi_0^\ast \rangle_R$ must have Chern number $-N$, therefore a fully dephased Chern insulator in the doubled space becomes the Gutzwiller projected QSH state~\cite{nayan2024,su2025,sarma2025}. Previous numerical results observed that, the Gutzwiller projected QSH insulator leads to quasi-long range inplane spin correlation~\cite{ran2008,sarma2025,wangbalents}, qualitatively consistent with our result that inplane pseudo-spin correlation upper-bounds other correlators. 

In the appendix we will discuss a version of ``Goldsteon theorem" in our set-up. The Goldstone theorem combined with Our inequality poses another strong constraint on the possible emergent order under general decoherence channels in Eq.~\ref{jump}. if there is an emergent order that spontaneously breaks the SU($N_f$) symmetry, it should not be caused by the condensate of a higher-fermion composite operator (composite of four or more fermions), without a condensate of SW-SSB order parameter in the form of ${\cal O} = c^\dagger \sigma^1 \Omega c$. 



\subsection{Dephased Dirac fermions}

In the previous section we proved the desired inequality for a lattice model. The same inequality can also be proven directly in the continuum, for example for $2d$ Dirac fermions, which may appear on the boundary of a $3d$ bulk rather than a lattice model in $2d$. We consider $N$-flavors of $2d$ Dirac fermions under dephasing. In the continuum the Dirac fermion Hamiltonian reads \beqn H_0 &=& \int d^2x \sum_{a = 1}^N \psi^\dagger_{a}  \left(  - \ii \tau^x \partial_x - \ii \tau^z \partial_y + m \tau^y \right) \psi_a.   
\eeqn  
We keep a finite $m$ for the discussions, but our inequality also holds in the limit of $m \ra 0$.
The parent Lagrangian for the doubled state $|\Psi(0)\rrangle $ reads: \beqn \cL_0 &=& \sum_{a = 1}^N \psi^\dagger_{a} D \psi_{a}, \cr\cr D &=& \partial_\tau  - (\ii \tau^x \partial_x + \ii \tau^z \partial_y + m \tau^y) \sigma^3. \eeqn Note that now $\psi$ carries the flavor, pseudo-spin, as well as the Dirac indices. $\tau^\mu$ and $\sigma^3$ act in the Dirac and pseudo-spin indices respectively. 

We evaluate the correlation function of the fermion-bilinear operator $O = \psi^\dagger \Gamma \Omega \psi$. 
$\Omega$ is a $N \times N$ Hermitian matrix in the flavor space, and $\Gamma$ is a matrix mixing both the Dirac and pseudo-spin index: $\Gamma = \tau^\mu \sigma^\nu$. 
In particular, the operator ${\cal O} = \psi^\dagger \sigma^1 \Omega \psi$ is a mass operator of the Dirac fermion, also a Cooper pair operator before transformation Eq.~\ref{PHR}. 
The correlator $C_\Gamma(x)$ of $O$ is evaluated as: \beqn C_\Gamma(x) &=& \frac{1}{\cZ} \int D \psi D \psi^\ast \ O_{0,\tau = 0} \, O_{x, \tau=0} \cr\cr &\times& \exp\left( - \int d\tau d^2x 
\ \cL_{0} - 2 H_{{\rm eff}}(g)_{\tau = 0} \right) . \label{PI} \eeqn 
The algebra Eq.~\ref{algebra1} again leads to the result that, the correlator $C_\Gamma(x)$ of $\psi^\dagger \Gamma \Omega \psi$ with any $\Gamma = \tau_\mu \sigma_\nu$ is upper-bounded by correlator ${\cal C}_\ast(x)$ of ${\cal O}$, for even integer $N$. 


In the massless limit, the initial doubled state has in total $2N$ flavors of $2d$ massless Dirac fermions, and in the strong dephasing limit, the dephased state in the doubled space becomes a Gutzwiller projected spin liquid state. It is often expected that this Gutzwiller construction (approximately) captures physics of a QED$_3$ spin liquid~\cite{hasting2000,hermele2005,ran2007,hermele2008,becca,becca2,hesl,hesl2,lauchlisl}, described by a field theory of $2N$ Dirac fermions coupled with a dynamical U(1) gauge field. The HS field $\phi$ corresponds to the temporal component of the gauge field. The $1/N$ calculation does suggest that the fermion bilinear operators with the form ${\cal O} = \psi^\dagger \sigma^1 \Omega \psi$ acquires a negative anomalous dimension, i.e. enhanced correlation from gauge fluctuation~\cite{hermele2005}, which is qualitatively consistent with our result. 

A similar inequality holds for QED$_3$ with $2N$ flavors of 2-component Dirac fermions and even integer $N$, with both the temporal and spatial components of the gauge field. The analogue of the key algebra Eq.~\ref{algebra1} is simply the fact that the gauged Dirac operator is anti-Hermitian: $\slashed{D}_a = - \slashed{D}_a^\dagger$, where $\slashed{D}_a = \gamma_\mu (\partial_\mu - \ii a_\mu)$. We also need the action of the gauge field to be a Maxwell theory without any Chern-Simons term. The inequality indicates that, in the massless limit, if QED$_3$ is a $(2+1)d$ conformal field theory, the anomalous dimension of operator ${\cal O} = \bar{\psi} \Omega \psi$ with nontrivial flavor matrix $\Omega$ cannot be positive. In other words gauge field fluctuation always enhances the correlation of ${\cal O}$. Let us first investigate another operator $\psi^\dagger \Omega \psi$, which is the density operator of a conserved charge of the $\SU(2N)$ flavor symmetry, therefore must have scaling dimension $2$ in QED$_3$. Then our inequality implies its correlator must be upper-bounded by the correlator of ${\cal O}$, meaning the anomalous dimension of ${\cal O}$ cannot be positive: \beqn \langle {\cal O}_0 \, {\cal O}_x \rangle \sim \frac{1}{|x|^{2\Delta}}, \ \ \ \Delta \leq 2, \eeqn which is consistent with previous $1/N$ calculation. 




\section{Summary}

In this work we discussed an inequality providing a bound for the behaviors of fermionic quantum matter under decoherence. Though the exact solution of the decohered quantum matters are unavailable in general, this inequality indicates that the system is driven towards SW-SSB under decoherence, as the R\'{e}nyi-2 correlator of a class of operators are upper-bounded by certain operators serving as a proximate diagnostic of SW-SSB. The proof is facilitated by the structure of the dephased quantum matter, particularly its natural form in the doubled-space representation.

In this work we only discussed the inequality for the R\'{e}nyi-2 correlators, thanks to its natural connection to Euclidean spacetime path-integral. Generalization of the inequality to fidelity or R\'{e}nyi-1 correlator, if possible, likely requires new techniques, and we leave this to future explorations.

The authors are supported by the Simons foundation through the Simons Investigator program. We thank Chong Wang for helpful discussions. We also acknowledge generative-AI for discussions and generating Fig.~\ref{diagram}.

\bibliography{refs}

\appendix

\section{A ``Goldstone" Theorem}

In this section we aim to prove a ``Goldstone" theorem on the temporal defect $\tau = 0$, and we use U(1) symmetry as an example: if an order parameter $O$ charged under a U(1) symmetry has long range correlation or a power-law correlation with sufficiently small scaling dimension at the temporal defect $\tau = 0$, then the U(1) current operator also cannot have short-range correlation on the temporal defect. Note that the notion of ``gapless" mode does not directly apply to our set up, therefore we only seek to address the equal-time correlation function. 

We apply the Ward identity on the temporal defect $\tau = 0$: \beqn && \partial_\mu \langle J_\mu(x,\tau) O(r, 0) O^\dagger(0, 0) \rangle \cr\cr &=& q\delta(\tau)(\delta^d(x-r) - \delta^d(x)) \langle O(r) O^\dagger(0) \rangle. \eeqn 
Here $q$ is the charge of $O$ under the U(1) symmetry. After performing Fourier transformation of $x_\mu$ in the equation above, and taking the frequency $k_0 = 0$, the Ward identity becomes \begin{equation}
 \sum_i k_i \cdot \langle J_i (k) O(r) O^\dagger(0) \rangle = q \langle O(r) O^\dagger(0) \rangle (e^{- \ii k \cdot r } - 1).
\end{equation} Here $q$ is the charge carried by the order parameter $O$.

We then use the Cauchy-Schwarz inequality \beqn |\langle J_i(k) X \rangle|^2 \leq \langle J_i(k) J_i(- k) \rangle \langle X^\dagger X \rangle , \eeqn and take $X = O( r ) O^\dagger( 0)$. We find
\beqn
    && |\langle J_i(k) O(r) O^\dagger(0) \rangle|^2 \cr\cr &\leq& \langle J_i(k) J_i(- k) \rangle \langle (O(r) O^\dagger(0))^\dagger O(r) O^\dagger(0) \rangle 
\eeqn

This inequality holds for any $r$. Let's just choose $k_x = k$, $k_y = 0$, $r_x = r$, $r_y = 0$. Combining the CS inequality and the Ward identity together, we have
\beqn
    &&  k^2 \langle J_x(k) J_x(- k) \rangle \cr\cr &\geq&  4q^2 \sin\left( k\cdot r /2 \right)^2  \frac{|\langle O(r) O^\dagger(0) \rangle|^2}{\langle (O(r) O^\dagger(0))^\dagger O(r) O^\dagger(0) \rangle}.
\eeqn
The denominator of the inequality above should saturate to a constant in the limit of $r \ra \infty$, as long as $\langle O^\dagger O \rangle$ is nonzero. 
\begin{equation}
    \langle J_x( k) J_x(- k) \rangle \gtrsim C q^2 \sin(k r/2)^2 \frac{1}{|r|^{4\Delta}} \frac{1}{k^2}
\end{equation}
We further choose $|r| = \pi/k$, then \begin{equation}
    \langle J_x( k) J_x(- k) \rangle \gtrsim C q^2 \frac{1}{k^{2 - 4\Delta}}.
\end{equation}

Therefore, for $\Delta < \frac{1}{2}$, this inequality forbids the current from having short-range correlations due to the singularity at small $k$. 

This Goldstone theorem helps constrain the possible emergent order under decoherence, combined with our inequality. 
If under decoherence there is an emergent order in the doubled space that spontaneously breaks the SU($N_f$) symmetry, it should {\it not} be caused by the condensate of a higher-fermion composite operator (composite of four or more fermions), without a condensate of SW-SSB order parameter in the form of ${\cal O} = c^\dagger \sigma^1 \Omega c$. The logic is that, if the continuous flavor symmetry is spontaneously broken, Goldstone theorem dictates that the symmetry current operator (a fermion bilinear) cannot be short-ranged. Then the inequality we proved implies that the correlator $\cC_\ast$ of ${\cal O}$ cannot be short-ranged, as $\cC_\ast$ bounds the current correlation. In fact, the Weingarten inequality in QCD was used to show that, within the assumptions made, the QCD cannot lead to higher operator condensate (such as tetraquark operator condensate) without the quark-bilinear operator condensate~\cite{koganinequality}.

\end{document}